\documentclass[a4paper]{jpconf}
\usepackage{graphicx}
\begin{document}
\title{Neutron Star Dynamics under Time Dependent External Torques}

\author{MA Alpar and E G\"{u}gercino\u{g}lu}

\address{Faculty of Engineering and Natural Sciences, Sabanc{\i} University, Orhanl{\i} -Tuzla, \.{I}stanbul 34956, Turkey}

\ead{alpar@sabanciuniv.edu}

\begin{abstract}
The two component model of neutron star dynamics describing the behaviour of the observed crust coupled to the superfluid interior has so far been applied to radio pulsars for which the external torques are constant on dynamical timescales. We recently solved this problem  under arbitrary time dependent external torques. Our solutions pertain to internal torques that are linear in the rotation rates, as well as to the extremely non-linear internal torques of the  vortex creep model. Two-component models with linear or nonlinear internal torques can now be applied to magnetars and to neutron stars in binary systems, with strong variability and timing noise. Time dependent external torques can be obtained from the observed spin-down (or spin-up) time series, $\dot{\Omega}(t)$. 
\end{abstract}
\section{The Two Component Model with Constant External Torques}

The two component model for neutron star dynamics was proposed immediately after the observation of the first pulsar glitches [1]. This model and its extension to nonlinear vortex creep coupling between the crust and the superfluid [2] have been applied with remarkable success to the postglitch relaxation data from radio pulsars, for which the external torques are constant on dynamical timescales. This talk presents, in summary, our recent solution of the two component models for linear and nonlinear coupling for the general case of arbitrary time dependent external torques [3]. These general solutions are applicable to the timing behaviour of neutron stars in X-ray binaries and magnetars, as well as timing noise from radio pulsars.   

Postglitch exponential relaxation  on timescales of days to weeks supported the theoretical expectations of superfluidity in the neutron star interior. As exponential relaxation is produced in linear systems, the original two component model of Baym et al [1] posits a linear coupling between the crust and the superfluid interior. The equations of the linear two component model are  
\begin{eqnarray}
I_{\rm c}\dot{\Omega}_{\rm c} + I_{\rm s}\dot{\Omega}_{\rm s} & = & N_{\rm ext} \\
\dot{\Omega}_{\rm s} = - \frac{\Omega_{\rm s} - \Omega_{\rm c}}{\tau_0} & = & - \frac{\omega}{\tau_0}.
\end{eqnarray} 
$I_{\rm c}$ and $I_{\rm s}$ denote the moments of inertia and $\Omega_{\rm c}$ 
and $\Omega_{\rm s}$ are the rotation rates of the crust and the superfluid, respectively. The rotational velocity lag between the two components is denoted 
by $\omega \equiv \Omega_{\rm s} - \Omega_{\rm c}$. The internal torque that the superfluid exerts on the crust is
\begin{equation}
N_{\rm int} \equiv -I_{\rm s}\dot{\Omega}_{\rm s}
\end{equation}
while $N_{\rm ext}$ denotes the external torque on the neutron star. The “crust” (normal matter) rotation rate $\Omega_{\rm c}$ obeys the Navier Stokes Equation, but is already in rigid body rotation on superfluid timescales. 
The superfluid rotation rate $\Omega_{\rm s}$ does not obey the Navier Stokes Equation. 
Physically, the superfluid spins down by a flow of quantized vortices away from the rotation axis. Under many kinds of physical interactions between the vortices and normal matter, the vortex current, and the superfluid rate are linear in the lag $\omega$.  

Under a constant external torque $N_{\rm ext}$ the system has a steady state,
\begin{eqnarray}
\dot{\omega}& = & 0 \nonumber \\
\dot{\Omega}_{\rm s} & = & \dot{\Omega}_{\rm c} = \frac{N_{\rm ext}}{I}. 
\end{eqnarray} 
The steady state value $\omega_{\infty}$ of the lag is determined by 
\begin{equation}
\frac{\omega_{\infty}}{\tau_0} = - \frac{N_{\rm ext}}{I}.
\end{equation}
The linear response to any glitch induced offset $\delta\omega(0)$ from steady state is exponential relaxation
\begin{equation}
\dot\Omega_{\rm c}= \dot\Omega_{\rm c}(0)-\frac{I_{\rm s}}{I}\frac{\delta\omega(0)}{\tau}e^{-t/\tau}.
\end{equation}
Several components i of exponential relaxation are indeed observed 
following pulsar glitches.
Analysis of postglitch relaxation yields $I_{\rm{s,i}}/I \sim 10^{-3} - 10^{-2}$ 
for each component of exponential relaxation. This points at the crust superfluid. 
With effective masses (entrainment) [4] taken into account, 
$I_{\rm{s,i}}/I < \sim 10^{-1}$,  pointing to the crust + outer core [5]. 
The core superfluid is already 
coupled tightly to the crust; the core is effectively a part of the crust, dynamically, 
on glitch and postglitch relaxation timescales [6].
Analysis in terms of the {\it two} component model is enough: Since 
$I_{\rm{s,i}}/I << 1$, the different superfluid components i with moments 
of inertia $I_{\rm{s,i}}$ can be handled with the crust in separate two 
component models and then the response of the crust to each can be superposed. 

\section{Vortex Creep and the Nonlinear Two Component Model}

A superfluid component with vortex pinning can spin down by the 
thermally activated flow (creep) of vortices against pinning potentials. 
The vortex creep model gives the spindown rate  
\begin{equation}
\dot\Omega_{\rm s}= -\frac{4\Omega_{\rm s}v_0}{r}\exp{\left(-\frac{E_{\rm p}}{kT}\right)}\sinh{\left(\frac{\omega}{\varpi}\right)} \equiv -f(\omega) ,
\end{equation}
where 
\begin{equation}
\varpi \equiv \frac{kT}{E_{\rm p}}\omega_{\rm cr},
\end{equation}
to replace Eq.(2). Here $E_{\rm p}$ is the pinning energy and $T$ is the temperature. The distance $r$ of the vortex lines from the rotation axis is approximately equal to the neutron star radius $R$ in the crust superfluid where creep takes place against the pinning sites in the crust lattice. The microscopic vortex velocity $v_{0}$ around pinning centres 
is $\approx 10^{7}$ cm/s [7]. The two component model is now nonlinear 
in the lag $\omega$.

The values of the parameters, in particular of $E_{\rm p}/kT$ can be such that Eq. (7)  requires $\sinh{(\omega/\varpi)}\gg 1$, so that $\sinh{(\omega/\varpi)} \simeq (1/2)\exp{(\omega/\varpi)}$. Equations (1) and (7) then yield
\begin{equation}
\dot\Omega_{\rm c}= \frac{N_{\rm ext}}{I_{\rm c}} + \frac{I_{\rm s}}{I_{\rm c}}\frac{\varpi}{2\tau_{\rm l}}\exp(\omega/\varpi).
\end{equation}
where
\begin{equation}
\tau_{\rm l} \equiv \frac{kT }{E_{\rm p}} \frac{R \omega_{\rm cr}}{4 \Omega_{\rm s} v_{0}} \exp \left( \frac{E_{\rm p}}{kT} \right). \nonumber
\end{equation}
For nonlinear coupling, the steady state value of the lag, $\omega_{\infty}$ is given by Eqs. (4) and (7).

The solution for the crust spindown rate under a constant external torque $N_{\rm ext} = I \dot\Omega_{\infty}$ is
\begin{equation}
\dot\Omega_{\rm c}(t)=\frac{I}{I_{\rm c}}\dot\Omega_{\infty}-\frac{I_{\rm s}}{I_{\rm c}}\dot\Omega_{\infty} \left[\frac{1}{1+\left[\exp{\left(\frac{t_0}{\tau_{\rm nl}}\right)}-1\right]\exp{\left(-\frac{t}{\tau_{\rm nl}}\frac{I}{I_{\rm c}}\right)}}\right],
\end{equation}
with a nonlinear creep relaxation time
\begin{equation}
\tau_{\rm nl}\equiv \frac{kT}{E_{\rm p}}\frac{I\omega_{\rm cr}}{N_{\rm ext}} =\frac{kT}{E_{\rm p}}\frac{\omega_{\rm cr}}{\vert\dot{\Omega}\vert_{\infty}},
\end{equation}
and recoupling (waiting) timescale
\begin{equation}
t_0 \equiv \frac{I\delta\omega}{N_{\rm ext}}=\frac{\delta\omega}{\vert \dot\Omega\vert_{\infty}}.
\end{equation}
Nonlinear creep regions are responsible 
for glitches through vortex unpinning avalanches leading to a 
decrease $\delta \Omega_{\rm s}(r)$ in those superfluid regions through which the unpinned vortices move rapidly during the glitch. Angular momentum conservation leads to 
$\delta\Omega_{\rm s} =(I_{\rm c}/I_{\rm s})\Delta \Omega_{\rm c}\gg\Delta\Omega_{\rm c}$, and $\delta\omega\cong\delta\Omega_{\rm s}$. Because of the very sensitive dependence of the creep rate on $\delta\omega$ this glitch induced offset from the steady state lag 
$\omega_{\infty}$ will stop creep altogether in the superfluid regions effected. As the same external torque is now acting on less moment of inertia, the observed spindown rate of the crust will suffer a glitch associated jump
\begin{eqnarray}
\frac{\Delta\dot{\Omega}_{\rm c}}{\dot{\Omega}_{\rm c}}\cong \frac{I_{\rm s}}{I_{\rm c}}.  
\end{eqnarray}
creep restarts after a 
waiting time $t_{0}\cong\delta\Omega_{\rm s}/\vert\dot{\Omega}\vert_{\infty}$. 
This "Fermi function" response was predicted as a standard signature of vortex creep response to a glitch involving vortex unpinning (see Fig. 4 in [2]) and later observed in the Vela pulsar [8]. The common, almost ubiquitous form of nonlinear response  is the constant second derivative $\ddot{\Omega}_{\rm c}$ interglitch 
timing behaviour observed in the Vela and other pulsars [9][10]. This is the response 
to a uniform density of vortices unpinned at the glitch leading to 
a range of waiting times throughout the superfluid regions. Such power law behaviour is characteristic of nonlinear dynamics. 

Two component models with linear or nonlinear internal torques have so far been applied to post-glitch or inter-glitch timing behaviour of radio pulsars [9][11][12][13][14], 
where the external torque, with a secular (characteristic) 
timescale $\tau_{\rm c} \equiv \Omega_{\rm c}/2|\dot{\Omega}_{\rm c}| \sim 10^3 - 10^6 $ yr is constant for timescales of observed postglitch relaxation.

While constant secular external torques are characteristic for radio pulsars, neutron stars in X-ray binaries, magnetars and transients exhibit strong variations in the observed spin-down or spin-up rates. This indicates variable external torques, including strong torque noise. 
\\
\\
\\
\\
\section{Linear Two-Component Model with a Time Dependent External Torque} 

The two component model postglitch response with a linear internal torque and a time varying external torque is described by 
\begin{equation}
\dot{\omega} = - \frac{\omega}{\tau} - \frac{N_{\rm ext}(t)}{I_{\rm c}}.
\end{equation}
The relaxation time $\tau$ is determined by the physics of the internal torque. 
Using an integrating factor $e^{t/\tau}$ the solution 
\begin{eqnarray}
\omega(t) & = & e^{-t/\tau}\left[\omega(0)-\frac{1}{I_{\rm c}}{\int_{0}^{t}e^{t'/\tau}}N_{\rm ext}(t')dt'\right], \nonumber \\
\dot\Omega_{\rm c}(t) & = & \frac{N_{\rm ext}(t)}{I_{\rm c}}+\frac{I_{\rm s}}{I}\left(\frac{e^{(-t/\tau)}}{\tau}\left[\omega(0)-\frac{1}{I_{\rm c}}\int_{0}^{t} e^{(t'/\tau)}N_{\rm ext}(t') d t'\right]\right).
\end{eqnarray}

\section{Vortex Creep Model with a Time Dependent External Torque} 

For vortex creep under a time dependent external torque Equations (1) and (7) lead to
\begin{equation}
\dot\omega=-\frac{I \varpi}{2I_{\rm c}\tau_{\rm l}}e^{\omega/\varpi}-\frac{N_{\rm ext}(t)}{I_{\rm c}}.
\end{equation}
In terms of $y \equiv \exp(-\omega/\varpi)$, Equation (17) becomes 
\begin{equation}
\frac{dy}{dt}-\frac{N_{\rm ext}(t)}{I_{\rm c}\varpi}y-\frac{I}{2I_{\rm c}\tau_{\rm l}}=0.
\end{equation}
This has an integration factor $\exp\left(-\frac{X(t)}{I_{\rm c}\varpi}\right)$ where 
\begin{equation}
X(t)=\int_{0}^{t}N_{\rm ext}(t') d t'.
\end{equation}
The solution for the angular velocity lag $\omega$ exhibits an exponential dependence on the postglitch 
lag $\omega(0)$ and on $X(t)$, the cumulative angular momentum transfer by the external torque:
\begin{eqnarray}
e^{-(\omega/\varpi)}& = & e^{-\omega(0)/\varpi}\exp\left(\frac{X(t)}{I_{\rm c}\varpi}\right)\nonumber \\
 & + & \exp\left(\frac{X(t)}{I_{\rm c}\varpi}\right)\int_{0}^{t}\frac{I}{2I_{\rm c}\tau_{\rm l}}\exp\left(-\frac{X(t')}{I_{\rm c}\varpi}\right)dt'.
\end{eqnarray}
From Equations (1), (17), and (20) the 
response of nonlinear creep to a glitch in the presence of a time dependent 
external torque is obtained:
\begin{eqnarray}
\dot\Omega_{\rm c}(t) & = & \frac{N_{\rm ext}(t)}{I_{\rm c}}\nonumber \\ 
& + & \frac{I_{\rm s}}{I_{\rm c}}\frac{\varpi}{2\tau_{\rm l}}\left[\frac{1}{\exp{\left(\frac{\frac{X(t)}{I_{\rm c}}-\omega(0)}{\varpi}\right)}+\exp{\left(\frac{X(t)}{I_{\rm c}\varpi}\right)}\int_{0}^{t}\frac{I}{2I_{\rm c}\tau_{\rm l}}\exp{\left(-\frac{X(t')}{I_{\rm c}\varpi}\right)}dt'} \right].
\end{eqnarray}

The nonlinear creep timescale and the waiting time are now time dependent, 
involving the running time average $<N_{\rm ext}(t)> \equiv X(t)/t$ of 
the external torque:
\begin{eqnarray}
\tau_{\rm nl}(t) & \equiv & \frac{kT}{E_{\rm p}}\frac{I\omega_{\rm cr}}
{<N_{\rm ext}(t)>}, \\
t_{0}(t) & \equiv & \frac{I\delta \omega(0) }{<N_{\rm ext}(t)>}.
\end{eqnarray}

\section{Applications }

Equations (16) and (21) show that the observed $\dot\Omega_{\rm c}(t)$ 
gives $N_{\rm ext}(t)/ I_{\rm c} $ to lowest order in $I_{\rm s}/I_{\rm c}< \sim 0.1$. 
The residuals after a first fit to the $\dot\Omega_{\rm c}(t)$ time series will be 
$I_{\rm s}/I_{\rm c}$ times a certain convolution of the external torque with the internal torque, as defined in Eq. (16) or Eq. (21), allowing for a consistency check of the model.

Let us consider three kinds of time dependent external torques; (i) an exponentially decaying external torque, (ii) power law time dependence and (iii) timing noise. 

(i) Take an exponentially decaying torque with time scale $\tau_{\rm d}$ added to the preglitch external torque $N_0$:
\begin{equation}
N_{\rm ext}(t)=N_0 + \delta N e^{-t/\tau_{\rm d}}=I\dot\Omega_{\infty} + \delta N e^{-t/\tau_{\rm d}}.
\end{equation}

For linear internal torques Equation (16) gives the solution
\begin{eqnarray}
\dot\Omega_{\rm c}(t)& = & \dot\Omega_{\infty}+\frac{\delta N}{I_{\rm c}}e^{-t/\tau_{\rm d}}\left[1-\frac{I_{\rm s}}{I}\frac{\tau_{\rm d}}{\tau_{\rm d}-\tau}\right]\nonumber \\ 
& + & \frac{I_{\rm s}}{I}\left[e^{-t/\tau}\left(\frac{\omega(0)}{\tau}+\frac{I}{I_{\rm c}}\dot\Omega_{\infty}+\frac{\delta N}{I_{\rm c}}\frac{\tau_{\rm d}}{\tau_{\rm d}-\tau}\right)\right].
\end{eqnarray}

For vortex creep the solution is obtained by substituting 
the integrated angular momentum transfer
\begin{equation}
X(t)=N_{0} t + \delta N \tau_{\rm d} [ 1 - e^{-t/\tau_{\rm d}}] 
\end{equation}
in Equation (21). 

(ii) For a power law  torque with index $\alpha$ added to the preglitch torque:
\begin{equation}
N_{\rm ext}(t)=N_0 + \frac{\delta N t_0^{\alpha}}{(t+t_0)^{\alpha}}=I\dot\Omega_{\infty} + \frac{\delta N t_0^{\alpha}}{(t+t_0)^{\alpha}}
\end{equation}
Equation (16) gives
\footnotesize
\begin{eqnarray}
\lefteqn{\dot\Omega_{\rm c}(t)=\dot\Omega_{\infty}+\frac{\delta N}{I_{\rm c}}\frac{t_0^{\alpha}}{(t+t_0)^{\alpha}}\left[1-\frac{I_{\rm s}}{I}\frac{(t+t_0)}{(1-\alpha)\tau}\right]{}}\nonumber \\
&&{}+\frac{I_{\rm s}}{I}\left[e^{-t/\tau}\left(\frac{\omega(0)}{\tau}+\frac{I}{I_{\rm c}}\dot\Omega_{\infty}+\frac{\delta N}{I_{\rm c}}\frac{t_0}{(1-\alpha)\tau}+\frac{I_{\rm s}}{I}\frac{\delta N}{I_{\rm c}}\frac{t_0^{\alpha}}{(1-\alpha)\tau^{2}}\int_{0}^{t} \frac{e^{t'/\tau}dt'}{(t+t_0)^{\alpha-1}}\right)\right],
\end{eqnarray}
\normalsize
for linear internal torques. 

The integrated angular momentum transfer $X(t) $ is 
\begin{equation}
X(t)=N_0 t + \frac{\delta N t_0}{\alpha - 1}\left[ 1 - \frac{t_0^{\alpha - 1}}{(t+t_0)^{\alpha - 1}}\right],
\end{equation}
defining the response with nonlinear internal torques through Equation (21). 

(iii) Take white torque noise: 
\begin{equation}
N_{\rm ext}=\sum_{\rm i}\alpha_{\rm i}\delta(t-t_{\rm i}),
\end{equation}
where $\alpha_{\rm i}$ are amplitudes of torque variations. 
For linear coupling Equations (16) and (30) lead to the power spectrum
\begin{eqnarray}
P(f) & = & \frac{1}{\sqrt{2\pi}}\left[\frac{2<\alpha^{2}>}{I_{\rm c}^{2}}+\left(\frac{I_{\rm s}/I}{1+(2\pi\tau f)^{2}}\right)\left(\frac{<\alpha><\omega>}{I_{\rm c}}-\frac{<\alpha^{2}>}{I_{\rm c}^{2}}\right)\right] \nonumber \\
& + & \frac{1}{\sqrt{2\pi}}\left[\left(\frac{(I_{\rm s}/I)^{2}}{1+(2\pi\tau f)^{2}}\right)\left(\frac{2<\alpha^{2}>}{I_{\rm c}^{2}}+2<\omega^{2}>-\frac{<\alpha><\omega>}{I_{\rm c}}\right) \right],
\end{eqnarray}
where $<\alpha>$ is the mean external torque variation amplitude and $<\omega>$ 
denotes the mean value of the angular velocity lag. This result was 
obtained earlier for the linear two component model [15][16][17]. 
To lowest order we find the model power spectrum, 
a constant $P(f)$ for white noise. To order $I_{\rm s}/I$ we find the 
power spectrum of the integrated process, which is flat at low $f$ and 
a random walk spectrum, $P(f)\propto f^{-2}$, at high 
frequencies $f\gg \tau^{-1}$. The first fit will give 
the strength of the noise process, the term proportional 
to $<\alpha^{2}>$. The residuals to order $I_{\rm s}/I$ will 
constrain $<\alpha><\omega>$ and $<\omega^{2}>$. The term 
proportional to $(I_{\rm s}/I)^{2}$ 
can be neglected to a good approximation. Noise processes other 
than white noise can be handled similarly for the linear two component model, by inserting the noise model in Eq.(16) and calculating the power spectrum. For the nonlinear two component model, Eq. (21) does not lead to explicit expressions for any time dependent external torque, but a well defined algorithm can be constructed to obtain the power spectrum, as for the times series in cases (i) and (ii).   

\subsection{Acknowledgments}
This work was supported by the Scientific and Technological Research Council of Turkey
(T\"{U}B\.{I}TAK) under the grant 113F354. M.A.A. is a member of the Science Academy
(Bilim Akademisi), Turkey.

\end{document}